\documentclass[showpacs,aps,amsmath,amssymb,twocolumn]{revtex4}
\usepackage[]{graphicx}			
\usepackage{color}
\newcommand{\be}{\begin{equation}}
\newcommand{\ee}{\end{equation}}
\newcommand{\ben}{\begin{eqnarray}}
\newcommand{\een}{\end{eqnarray}}
\newcommand{\bes}{\begin{subequations}}
\newcommand{\ees}{\end{subequations}}
\newcommand{\bb}{\bibitem}

\newcommand{\nn}{\nonumber\\}
\newcommand{\bfi}{\begin{figure}}
\newcommand{\efi}{\end{figure}}
\newcommand{\bc}{\begin{center}}
\newcommand{\ec}{\end{center}}
\newcommand{\sech}{\mbox{sech}}
\newcommand{\arcsinh}{\mbox{arcsinh}} 
\begin{document}
\title{Thick brane models in generalized theories of gravity}
\author{D. Bazeia$^{1}$, A. S. Lob\~ao Jr.$^{2}$ and R. Menezes$^{3,4}$}
\affiliation{$^1$Departamento de F\'\i sica, Universidade Federal da Para\'\i ba, 58051-970 Jo\~ao Pessoa, PB, Brazil\\ 
$^2${Escola T\'ecnica de Sa\'ude de Cajazeiras, Universidade Federal de Campina Grande, 58900-000 Cajazeiras, PB, Brazil}\\
$^3$Departamento de Ci\^encias Exatas, Universidade Federal da Para\'\i ba, 58297-000 Rio Tinto, PB, Brazil and
\\$^4$Departamento de F\'\i sica, Universidade Federal de Campina Grande, 58109-970 Campina Grande, PB, Brazil.}
\begin{abstract}
This work deals with thick braneworld models, in an environment where the Ricci scalar is changed to accommodate the addition of two extra terms, one depending on the Ricci scalar itself, and the other, which takes into account the trace of the energy-momentum tensor of the scalar field that sources the braneworld scenario. We suppose that the scalar field engenders standard kinematics, and we show explicitly that the gravity sector of this new braneworld scenario is linearly stable. We illustrate the general results investigating two distinct models, focusing on how the brane profile is changed in the modified theories.
\end{abstract}
\maketitle

\section{introduction}

This work deals with braneworld models in the presence of scalar fields \cite{RS,GW,F} in an AdS$_5$ geometry with a single extra dimension of infinite extent. Differently from the original Randall and Sundrum (RS) proposal presented in \cite{RS}, which leads to the thin braneworld scenario, the coupling with scalar fields is also of interest and leads to another, thick braneworld scenario \cite{GW,F,C}. The presence of a scalar field makes the warp function to behave smoothly, leading to a thick braneworld scenario. This possibility has opened a new area of study, and here we quote Ref.~\cite{G} for some works on the subject.

In the current work, we go beyond General Relativity. Before doing this, however, we should care about issues concerning the presence of degrees of freedom which may lead to instabilities in these theories, and here we recommend the interesting reviews on the subject \cite{ghost}. To go beyond General Relativity, we add two distinct contributions, one taking the metric tensor $g_{\mu\nu}$ to change $R=g_{\mu\nu}R^{\mu\nu}$ (the Ricci scalar, the trace of the Ricci tensor) to $F(R)$, and the other, modifying Einstein's equation with the inclusion of auxiliary fields, as suggested recently in Ref.~\cite{psv}. To be more specific, we recall that the modification that changes $R$ into $F(R)$ in the Einstein-Hilbert action is older and was reviewed in Ref.~\cite{FR}. The modification introduced in the presence of auxiliary fields is more recent \cite{psv} and was studied in Refs.~\cite{c1,c2} within the cosmological context, to see how the auxiliary fields may contribute to the cosmic evolution. Moreover, it has also been recently studied within the thick braneworld context, in five dimensions with a single extra dimension of infinite extent \cite{gly,we}. See also Refs.~\cite{m1,m2} for other related studies on thick branes.

In order to widen the interest in the work, here we add together both $R$ and $T=g_{\mu\nu}T^{\mu\nu}$ (the trace of the energy-momentum tensor) into $F(R,T)$ to describe the generalized model, inspired by the recent investigations \cite{RT1,RT2}. We then study how the two modifications act to change the standard braneworld scenario, in an $AdS_5$ braneworld geometry with a single extra spatial dimension of infinite extent. We implement the investigation in Sec.~\ref{sec:gen}, where we discuss the general theory, including linear stability of the gravity sector. We then study two specific new models in Sec.~\ref{sec:exe}, working out the most important results of the models. As one knows, scalar fields play a fundamental role in cosmology as possible explanations for inflation, late time acceleration, and dark matter, among other issues. Also, scalar fields are fundamental sources in the thick braneworld scenario with a single extra dimension of infinite extent. Thus, here we investigate scenarios described by a single real scalar field to source the thick brane configuration, leaving aside studies related to cosmology. We then end the work in Sec.~\ref{sec:end}, where we include our comments and conclusions.

\section{The problem}\label{sec:gen}

We start investigating a model that describes generalized gravity coupled to a scalar field in five-dimensional space-time, with a single extra dimension $y$ of infinite extent. The model is controlled by an action of the form
\begin{equation}\label{eq1}
S=\int d^4x\,dy \sqrt{|g|}\Big(-\frac14F(R,T)+{\cal L}_s\Big)\,,
\end{equation}
where $R$ and $T$ represents the Ricci scalar and the trace of the energy-momentum tensor, and ${\cal L}_s$ describes the source Lagrange density, which we suppose is of the form
\begin{equation}\label{eq2}
{\cal L}_s=\frac12\nabla_a\phi\nabla^a\phi-V(\phi)\,.
\end{equation}
It engenders standard kinematics, and we are using $4\pi G^{(5)}=1$, $g=det(g_{ab})$, $a,b=0,1,...,4$ and the signature of the metric is $(+----)$. We focus attention on the profile of the thick braneworld scenario, so the addition of fermions and gauge fields will not be considered in this work. We take the spacetime coordinates and fields as dimensionless quantities, and we use Greek indices for the embedded (3+1)-dimensional space, $\mu, \nu =0,1,2,3$. 

We can use the source Lagrange density \eqref{eq2} to write the energy-momentum tensor in the form
\begin{equation}\label{eq3}
T_{ab}=-\frac12g_{ab}\nabla_c\phi\nabla^c\phi+g_{ab}V+\nabla_a\phi\nabla_b\phi\,,
\end{equation}
which contributes to give the trace
\be
T=-(3/2)\nabla_a\phi\nabla^a\phi+5V.
\ee
Moreover, the equation of motion for the scalar field has the form
\ben\label{eq4}
\nabla_a\nabla^a\phi+\frac34\nabla_a\Big(F_T \nabla^a\phi\Big) +\frac54V_\phi F_T+ V_\phi=0,
\een
where $V_\phi=dV/d\phi$, and $F_R=dF/dR$, $F_{T}=dF/dT$, etc. By varying the action of the gravitational field with respect to the metric tensor we obtain the modified Einstein equation
\ben
F_R R_{ab}-\frac12 g_{ab}F+\Big(g_{ab}\Box-\nabla_a\nabla_b\Big)F_R\nn=
2T_{ab}+\frac32 F_T\nabla_a\phi\nabla_b\phi\,.\label{eq5}
\een

We study the case of a flat brane, with the line element
\begin{equation}\label{eq7}
ds^2=e^{2A}\eta_{\mu\nu}d x^\mu d x^\nu-dy^2\,,
\end{equation}
where $A$ is the warp function, $e^{2A}$ is the warp factor and $\eta_{\mu\nu}$ is the 4-dimensional Minkowski metric, with signature $(+,-,-,-)$. We consider the case where both $A$ and $\phi$ are static and only depends on the extra dimension, that is, $A=A(y)$ and $\phi=\phi(y)$. We then get
\begin{equation}\label{eq8}
\Big(1+\frac34 F_T\Big)\phi^{\prime\prime}+\Big[(4+3F_T)A^\prime+\frac34 F_T^\prime\Big]\phi^\prime=\Big(1+\frac54 F_T\Big)V_\phi\,,
\end{equation}
where the prime denotes derivative with respect to the extra dimension. Also, the modified Einstein equation becomes
\bes\label{eq9}
\ben
\frac23\phi^{\prime 2}\Big(\!1\!+\!\frac34 F_T\!\Big)\!\!\!&=&\!\!\!-A^{\prime \prime} F_R +\frac13A^\prime F_R^\prime-\frac13F_R^{\prime\prime}\,,\label{eq9a}\\
\!\!\!\!\!\!V(\phi)\!-\!\frac{\phi^{\prime 2}}2\Big(\!1\!+\!\frac32 F_T\!\Big)\!\!\!&=&\!\!\!2(A^{\prime 2}\!\!+\!A^{\prime \prime})F_R\!-\!\frac{F}4\!-\!2A^\prime F_R^\prime\label{eq9b}.
\een
\ees
As it is standard check, here we also note that the set of equations \eqref{eq9} leads us with the equation of motion \eqref{eq8}. 

An important characteristic of the brane is its tension, which is given by
\begin{equation}
{\cal T}=\int dy\,\rho(y)\,,
\end{equation}
where $\rho(y)=-e^{2A}{\cal L}_s$ is the energy density of the brane. 

The next important step is to check if the modification proposed above contribute to destabilize the geometric degrees of freedom of the braneworld model. We investigate this issue studying linear stability of the gravity sector in the usual way.

\subsection{Stability}

The investigation of linear stability of the braneworld model can be done assuming that the metric is perturbed in the form
\be\label{eq10}
ds^2=e^{2A(y)}\Big[\eta_{\mu\nu}+h_{\mu\nu}(y,x)\Big] dx^\mu dx^\nu -dy^2\,.
\ee
Furthermore, the scalar field is written in the form
\be\label{eq11}
\phi=\phi(y)+\xi(y,x)\,.
\ee
The first-order contributions in $h$ and $\xi$ that follow from the equations of motion lead us to
\ben
&&\Big[\frac12h_{\mu\nu}^{\prime\prime}+2A^\prime h_{\mu\nu}^\prime+\frac12\eta_{\mu\nu}A^\prime h^\prime-\frac12e^{-2A}\Box^{(4)} h_{\mu\nu}+\nn
&&+\frac12e^{-2A}\Big(\partial_\mu\partial^\alpha h_{\alpha\nu}+\partial_\nu\partial^\alpha h_{\alpha\mu}-\partial_\mu\partial_\nu h\Big)\Big]F_R -
\nn
&&-e^{-2A}\partial_\mu\partial_\nu P(x,y)+\frac12h_{\mu\nu}^\prime F_R^\prime= Q(x,y)\eta_{\mu\nu} \,,\label{eq12}
\een
where $h=h^{\mu}{}_\mu$. For simplicity, we have introduced the functions $P(x,y)$ and $Q(x,y)$, which are defined as
\ben
P(x,y)\!\!&=&\!\!\Big[h^{\prime\prime}+5A^\prime h^\prime+e^{-2A}\Big(\partial^\mu\partial^\nu h_{\mu\nu}-\Box^{(4)}h\Big)\Big]F_{RR} \nn
\!\!&&\!\!+\Big[5V_\phi\xi+3\phi^\prime\xi^\prime+(\phi^{\prime 2}+2V)h\Big]F_{RT}\,,\label{eq13}
\een
and
\ben
Q(x,y)\!\!&=&\!\!3A^\prime P^\prime(x,y)+P^{\prime\prime}(x,y)-e^{-2A}\Box^{(4)} P(x,y)+\nn
\!\!&&\!\!+\Big(A^\prime h+\frac12 h^\prime\Big)F_R^\prime+2V_\phi\xi+2\phi^\prime\xi^\prime+\Big[\frac12F_T-\nn
\!\!&&\!\!-F_{RT}(4A^{\prime 2}\!+\!A^{\prime\prime})\!\Big] \!\Big[5V_\phi\xi\!+\!3\phi^\prime\xi^\prime\!\!+\!(\phi^{\prime 2}\!\!+\!2V)h\Big]\!\!-\nn
\!\!&&\!\!-\Big[(4A^{\prime 2}+A^{\prime\prime})F_{RR}-\frac12F_R\Big]\Big[h^{\prime\prime}+5A^\prime h^\prime+\nn
\!\!&&\!\!+e^{-2A}\Big(\partial^\mu\partial^\nu h_{\mu\nu}-\Box^{(4)}h\Big)\Big]\,.\label{eq14}
\een
If we take $F(R,T)\to R$, in the case of General Relativity we get $P(x,y)=0$ and
\ben
Q(x,y)&=&2V_\phi\xi+2\phi^\prime\xi^\prime+\frac12h^{\prime\prime}+\frac52A^\prime h^\prime\nonumber\\
&&+\frac12e^{-2A}\Big(\partial^\mu\partial^\nu h_{\mu\nu}-\Box^{(4)}h\Big).
\een
We can simplify the investigation of stability considering the transverse traceless components of metric fluctuations, that is, we take
\begin{equation}\label{eq15}
\partial^\mu h_{\mu\nu}=0;\;\;\;\;\;h=0\,.
\end{equation}
Thus, we can check that Eq.~\eqref{eq12} reduces to the form
\ben
&&\Big(\!-\partial_y^2 -4A^\prime \partial_y+e^{-2A}\Box^{(4)} -\frac{F_R^\prime}{F_R}\partial_y\Big)h_{\mu\nu}=\nn 
&&\frac12e^{-2A}\frac{F_{RT}}{F_R}\Big[\eta_{\mu\nu}\Box^{(4)}\!-\!4\partial_\mu\partial_\nu \Big](5V_\phi\xi\!+\!3\phi^\prime\xi^\prime) \,.\label{eq16}
\een
We see that the analysis of stability depends crucially on the function $F(R,T)$, as expected. The problem simplifies if we assume that the function $F(R,T)$ is separable in the form: $F(R,T)=G(R)+H(T)$. This choice allows to write the previous equation as
\ben
&&\Big(\!-\!\partial_y^2\! -\!4A^\prime \partial_y\!+\!e^{-2A}\Box^{(4)} \!-\!\frac{G_R^\prime}{G_R}\partial_y\!\Big)h_{\mu\nu}=0 \,.\label{eq17}
\een

We introduce the $z$-coordinate in order to make the metric conformally flat. We take $dz=e^{-A(y)}dy$ and we write
\ben\label{eq18}
h_{\mu\nu}(x,z)=e^{-ip\cdot x}e^{-3A(z)/2 }G_R^{-1/2}\bar h_{\mu\nu}(z) \,.
\een
In this case, the 4-dimensional components of $h_{\mu\nu}$ obey the Klein-Gordon equation and the metric fluctuations of the brane solution lead to the Schroedinger-like equation
\ben\label{eq19}
\left(-\frac{d^2}{dz^2} + U(z) \right)\bar h_{\mu\nu} = p^2 \bar h_{\mu\nu}\,,
\een
where 
\ben\label{eq20}
U(z)&=&\frac{9}{4} A_z^2 + \frac32 A_{zz}+\frac32A_z \frac{d(\ln G_R)}{dz}-\nn
&&-\frac14 \Big(\frac{d(\ln G_R)}{dz}\Big)^2+\frac1{2G_R}\frac{d^2 G_R}{dz^2}\,.
\een
Fortunately, we can use this $U(z)$ to write
\be
S^\dagger S=-\frac{d^2}{dz^2} + U(z)\,,\label{eq21}
\ee
where
\be
S=-\frac{d}{dz}+\frac32 A_z+\frac12 \frac{d(\ln G_R)}{dz}\,.\label{eq22}
\ee
The operator $-\partial_z^2 + U(z)$ is then non-negative, and so the gravity sector is linearly stable. This is an interesting result, which shows that despite the modification introduced with $F(R,T)$, when we write $F(R,T)=G(R)+ H(T)$, the gravity sector of the braneworld is linearly stable. We note that the stability behavior of the gravity sector only depends on the warp function, since $R$ only depends on $A(z)$ in this case.

\section{Specific models}\label{sec:exe}

Up to here, we used the generalized $F(R,T)$ gravity coupled to a real scalar field to obtain the equations of motion and show that the gravity sector of the thick braneworld scenario is linearly stable for $F(R,T)=G(R)+H(T)$. The results motivate us to further explore specific models, investigating how modifications depending on $R$ and $T$ change the standard scenario. We do this below, studying two distinct models.

\subsection{A simple model}

The first example we consider is described by the $F(R,T)$ function
\begin{equation}\label{eq23}
F(R,T)=R-\alpha\, T^{\,n},
\end{equation}
where $\alpha$ and $n$ are real parameters. This is a simple choice, and now the two components of equation \eqref{eq9} become
\bes\label{eq24}
\ben
\!\!\!\!\!A^{\prime \prime}\!\!\!&=&\!\!\!-\frac23\phi^{\prime 2}+ \frac{\alpha n }2 ~ \phi^{\prime 2}T^{n-1}\,,\label{eq24a}\\
\!\!\!\!\!A^{\prime 2}\!\!\!&=&\!\!\!\frac16\phi^{\prime 2}-\frac13V- \frac{\alpha n }4\phi^{\prime 2}T^{n-1}+\frac{\alpha}{12} T^n. \label{eq24b}
\een
\ees
where $T=\frac32 \phi^{\prime 2}+5V$. 

We concentrate on the simplest case, with $n=1$. Here the two components of the Einstein equation becomes
\bes\label{eq26}
\ben
\!\!\!\!\!A^{\prime \prime}\!\!\!&=&\!\!\!- \frac{4-3\alpha }6 ~ \phi^{\prime 2}\,,\label{eq26a}\\
\!\!\!\!\!A^{\prime 2}\!\!\!&=&\!\!\!\frac{4-3\alpha}{24}\phi^{\prime 2}-\frac{4-5\alpha}{12}V\,. \label{eq26b}
\een
\ees
To get to the first-order formalism, we introduce another function, $W = W(\phi)$, which can be used to see the warp factor as a function of the scalar field. We do this writing the first-order equation
\begin{equation}\label{eq27}
A^\prime=-\frac\gamma 3 W\,,
\end{equation}
where $\gamma$ is a real parameter to be determined. With this, we obtain that
\begin{equation}\label{eq28}
\phi^{\prime}=\frac{2\gamma}{4\!-\!3\alpha}W_\phi\,.
\end{equation}
Note that if $\gamma=1-3\alpha/4$ the solution is identical to the standard case ($\alpha=0$). On the other hand, for a general $\gamma$ the solutions depend on the parameter $\alpha$. Using equations \eqref{eq27} and \eqref{eq28} we can also obtain the potential as
\begin{equation}\label{eq29}
V(\phi)=\frac{2\gamma^2 ~W_\phi^2}{(4\!-\!5\alpha)(4\!-\!3\alpha)}-\frac{4\gamma^2W^2}{3(4\!-\!5\alpha)}\,.
\end{equation}
Furthermore, the energy density can be written in the form
\begin{equation}\label{eq31}
\rho(y)=\frac{4\gamma^2e^{2A(y)}}{4\!-\!5\alpha}\Big[\frac{4\!-\!4\alpha}{(4\!-\!3\alpha)^2}~W_\phi^2(\phi(y))-\frac{1}{3}W^2(\phi(y))\Big]\,.
\end{equation}
The energy of the thick brane becomes
\begin{equation}\label{eq32}
E_\alpha=\frac{\alpha \gamma^2}{(4\!-\!5\alpha)}\frac{8\!-\!9\alpha}{(4\!-\!3\alpha)^2}~\int_{-\infty}^{\infty} \!dy~ e^{2A(y)} W_\phi^2(\phi(y))\,.
\end{equation}
We note that the energy is controlled by $\alpha$, and for $0<\alpha<4/5$ the energy of the brane is positive.

\begin{figure}[t]
\begin{center}
\includegraphics[scale=0.7]{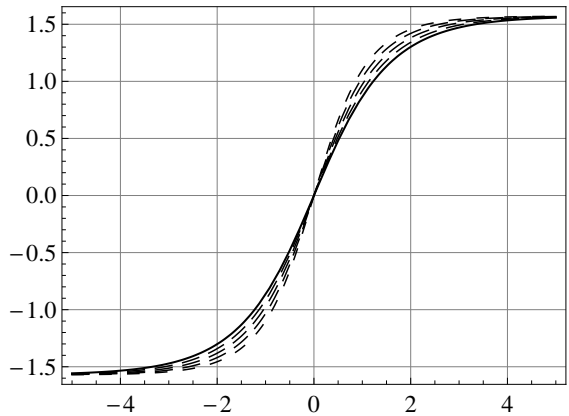}
\end{center}
\vspace{-0.7cm}
\caption{\small{The solution \eqref{eq33}, depicted for $a=b=\gamma=1$ and $\alpha\!=\!0$ (solid line), $\alpha\!=\!0.1, 0.2, 0.3, 0.4$ (dashed lines).\label{fig1}}}
\end{figure}

An interesting example which leads to analytical investigation of its stability is given with the superpotential in the form 
\begin{equation}\label{eq32}
W(\phi)= 2a \sin(b\phi)\,,
\end{equation}
where $a$ and $b$ are real parameters. With this choice, we obtain the solution \eqref{eq28} as
\begin{equation}\label{eq33}
\phi(y)=\frac1b\arcsin\Big[\tanh (By)\Big]\,,
\end{equation}
where $B\equiv 4\gamma ab^2/(4-3\alpha)$. In Fig.~\ref{fig1} we depict the behavior of the solution \eqref{eq33}. We note that the thickness of the solution is controlled essentially by $\alpha$. Moreover, as expected, the solution goes to the asymptotic value $ \phi\to \bar \phi=\pm \pi/2b$ when $y\to\pm\infty$.

\begin{figure}[t]
\begin{center}
\includegraphics[scale=0.7]{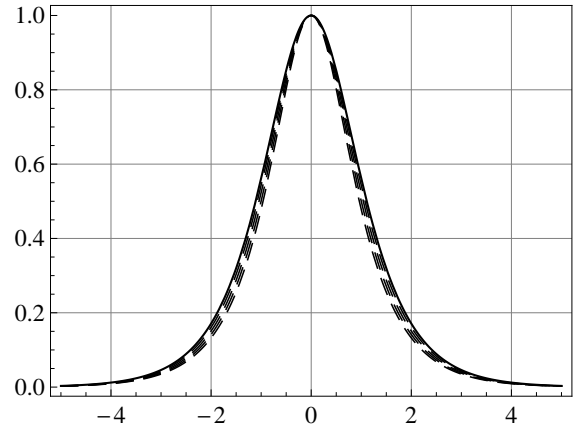}
\end{center}
\vspace{-0.7cm}
\caption{\small{The warp factor $e^{2A}$, depicted for $a=b=\gamma=1$ and $\alpha\!=\!0$ (solid line), and for $\alpha\!=\!0.1, 0.2, 0.3, 0.4$ (dashed lines).\label{fig2}}}
\end{figure}

Using equation \eqref{eq27} we can also get the warp function as
\begin{equation}\label{eq34}
A(y)=\frac{2 \gamma a}{3 B}~\ln \Big[\sech ( B y)\Big]\,.
\end{equation}
If $\alpha<4/3$ we can get the thin-brane limit when $y\to\pm\infty$, in the form
\begin{equation}\label{eq35}
A(y\to\pm\infty)\backsim -\frac{2 \gamma a}{3 } |y|\,.
\end{equation}
In Fig.~\ref{fig2} we depict the warp factor $e^{2A}$ for some values of the parameters. We note that the warp factor narrows as $\alpha$ increases.

\begin{figure}[t]
\begin{center}
\includegraphics[scale=0.7]{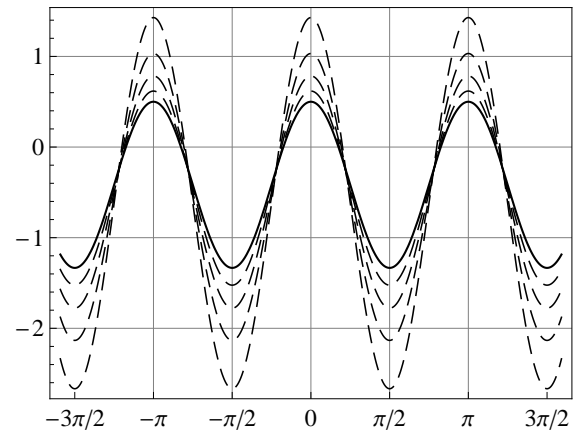}
\end{center}
\vspace{-0.7cm}
\caption{\small{The scalar field potential, depicted for $a=b=\gamma=1$ and $\alpha\!=\!0$ (solid line), $\alpha\!=\!0.1, 0.2, 0.3, 0.4$ (dashed lines).\label{fig3}}}
\end{figure}

Furthermore, the potential \eqref{eq29} can be written as
\begin{equation}\label{eq36}
V(\phi)=-\frac{16(a\gamma)^2}{3(4\!-\!5\alpha)}+\frac{2\gamma a(3B+8\gamma a)}{3(4\!-\!5\alpha)}\cos^2(b\phi)\,,
\end{equation}
which is depicted in Fig. \ref{fig3}. Note that the potential has several global minima at $\phi=\bar\phi_i$, where
\begin{equation}\label{eq37}
V(\bar \phi_i)=-\frac{16}{3}\frac{ (a\gamma)^2}{ (4-5 \alpha)}\,.
\end{equation}

\begin{figure}[t]
\begin{center}
\includegraphics[scale=0.7]{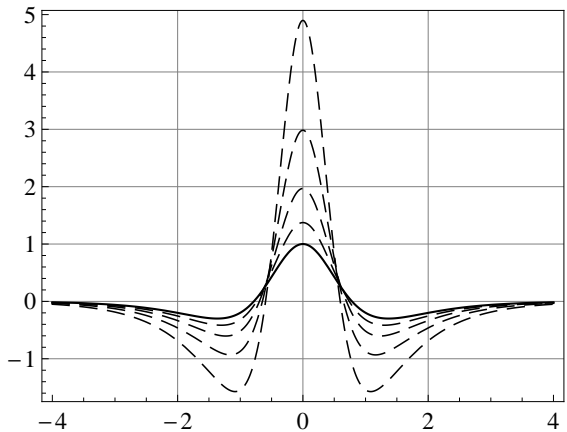}
\end{center}
\vspace{-0.7cm}
\caption{\small{The energy density, depicted for $a=b=\gamma=1$ and $\alpha\!=\!0$ (solid line), and for $\alpha\!=\!0.1, 0.2, 0.3, 0.4$ (dashed lines).\label{fig4}}}
\end{figure}

Finally, the energy density \eqref{eq31} can be written as
\ben\label{eq38}
\rho(y)\!\!\!&=&\!\!\!\frac{16(\gamma a)^2}{3(4\!-\!5\alpha)}\!\Big[\!\!-\!1\!+\!\!\Big(\!1\!+\!\frac{3B(1\!-\!\alpha)}{\gamma a(4\!-\!3\alpha)}\!\Big)\sech^2(B y)\!\Big]\!\!\times\nn
\!\!\!&&\!\!\!  \times \Big[\sech (B y) \Big]^{\frac{4\gamma a}{3B}}\,.
\een
Fig.~\ref{fig4} shows how the parameter $\alpha$ controls the energy density. The equation \eqref{eq38} can be integrated to give the energy of the brane in the form
\begin{equation}\label{eq39}
E_\alpha=\alpha \frac{B}{4b^2}\Big(\frac{8\!-\!9\alpha}{4\!-\!5\alpha}\Big)~\frac{ \sqrt{\pi}~\Gamma \left(s\right)}{\Gamma (s+1/2)}\,,
\end{equation}
where $s\equiv 1+2\gamma a/3B$.

\begin{figure}[t]
\begin{center}
\includegraphics[scale=0.7]{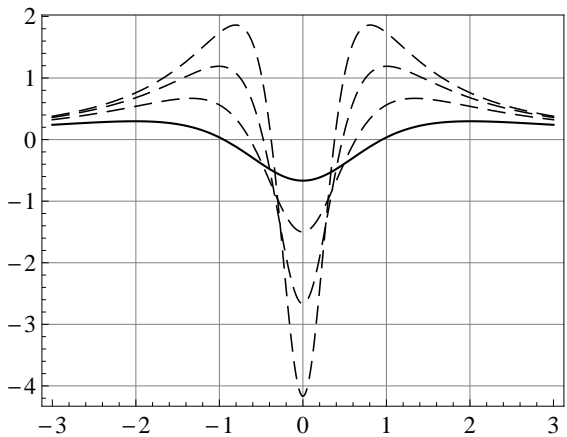}
\end{center}
\vspace{-0.7cm}
\caption{\small{The stability potential, depicted for $B\!=\!2/3$ (solid line), and for $B\!=\!1, 4/3, 5/3$ (dashed lines).\label{fig5}}}
\end{figure}

The behavior of the  stability of the gravitational sector can be analyzed analytically for the particular case where $B=2 \gamma a/3$. In this case we can make the change
\begin{equation}\label{eq40}
y=\frac1B \arcsinh(Bz)\,.
\end{equation}
With this we get the stability potential in the form
\begin{equation}\label{eq41}
U(z)=-\frac{3 B^2}{4}\frac{2-5B^2 z^2 }{ \left(1+B^2 z^2\right)^{2}}\,.
\end{equation}
It is depicted in Fig.~\ref{fig5}, for several values of the parameters.

The above results show that the thick brane narrows as $\alpha$ increases, but the brane profile is qualitatively similar to the case of a standard brane.

\subsection{Another model}

\begin{figure}[t]
\begin{center}
\includegraphics[scale=0.7]{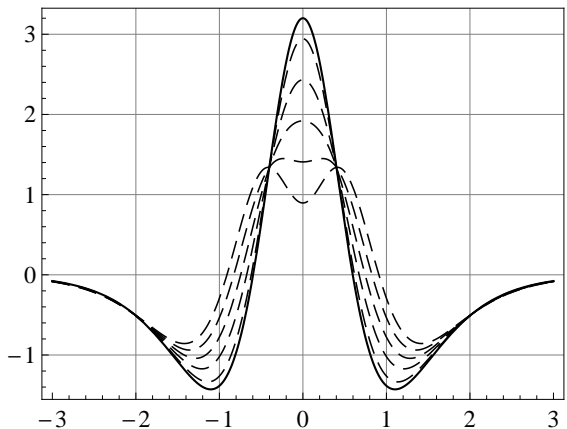}
\end{center}
\vspace{-0.7cm}
\caption{\small{The energy density, depicted for $k=1$, $\alpha=1/2$ and $\beta\!=\!0$ (solid line), $\beta\!=\!0.002, 0.006, 0.010, 0.014, 0.018$ (dashed lines).\label{fig6}}}
\end{figure}

Let us go on and investigate another model, this time with contribution from the Ricci scalar too. We assume that
\begin{equation}\label{eq42}
F(R,T)=R+\beta R^2-\alpha T\,,
\end{equation}
where $\alpha$ and $\beta$ are real parameters. The case with $\alpha=0$ has been studied in \cite{Bazeia:2013uva}, and now we add both the $R^2$ and $T$ contributions, to get analytical results. In Ref.~\cite{Bazeia:2013uva} one noted that the term $R^2$ contributes to generate an interesting effect, the splitting of the brane, and here we want to see how it works. In this new model, the brane equations become
\bes\label{eq43}
\ben
\!-\frac{4\!-\!3\alpha}{6} \phi^{\prime 2}\! \!=\!\!
 A^{\prime \prime}\!\!+\!\frac{8\beta}3\Big(\!2 A^{\prime \prime \prime \prime }\!\!+\!16 A^{\prime\prime 2}\!\!+\!8A^\prime \!\!A^{\prime \prime \prime } \!\!+\!5 A^{\prime 2}\!\! A^{\prime \prime }\!\Big),\nn\label{eq43a}\\
\frac{4-3\alpha}{24} \phi^{\prime 2}-\frac{4-5\alpha}{12} V=
A^{\prime 2}+\quad\quad\quad\quad\quad\quad\quad~~~~\nn
+\frac{4\beta}3\Big(5  A^{\prime 4}\!\!+8 A^{\prime }A^{\prime \prime \prime}\!\!+32  A^{\prime 2 }A^{\prime \prime }\!\!-4  A^{\prime \prime 2}\Big).~~~\label{eq43b}
\een
\ees
We note that if $\beta=0$ we get to the equations \eqref{eq26}. Also, for $\beta\neq 0$ the equations involve the fourth derivative of the warp function, which makes it difficult to search for solutions. However, as suggested in \cite{Bazeia:2013uva} if we choose the warp function $A(y)$ as
\begin{equation}\label{eq44}
A(y)=\ln \Big[\sech(k y)\Big]\,,
\end{equation}
where $k$ is a positive parameter, this allows us to write the relations \eqref{eq43} as
\bes \label{eq45}
\ben
\phi^{\prime2}\!\!\!&=&\!\!\!\frac{6k^2}{4-3\alpha} S^2\!+\!\frac{16\beta k^4}{4-3\alpha}  S^2\Big(29-49S^2\Big)\,, \label{eq45.a}\\
V(y)\!\!\!&=&\!\!\!-\frac{12k^2}{4\!-\!5\alpha}\!+\!\frac{15k^2}{4\!-\!5\alpha}S^2\!-\!\frac{8\beta k^4}{4\!-\!5\alpha} \Big(\!10\!-\!145S^2\!\!+\!147 S^4 \!\Big)\,,\nn\label{eq45.b}
\een
\ees 
where $S=\sech(ky)$. We use equation \eqref{eq45.a} to see that $\beta$ should be limited to the interval
\begin{equation}\label{eq46}
-\frac{3}{232k^2}\leq \beta\leq \frac{3}{160k^2}\,.
\end{equation}
Moreover, we suppose that the parameter $\alpha$ is positive, smaller than $4/5$. For this model, we can write the energy density as
\ben\label{eq47}
\rho(y) \!\!&=&\!\!-\frac{12k^2}{4\!-\!5\alpha}S^2\!+\!\frac{12k^2(6\!-\!5\alpha)}{(4\!-\!3\alpha)(4\!-\!5\alpha)}S^4\!-\!\frac{8\beta k^4}{4\!-\!5\alpha} S^2\Big[10-\nn
\!\!&&\!-\frac{116(6-5\alpha)}{4\!-\!3\alpha}S^2\!+\!\frac{98(8-7\alpha)}{4\!-\!3\alpha}S^4\Big]\,.
\een 
A closer examination of the energy density \eqref{eq47} shows that it start to split when
\begin{equation}\label{eq48}
\beta= \frac{3 (8-7 \alpha)}{16 (125-116 \alpha)k^2}\,.
\end{equation}
The splitting depends on $\alpha$, so the modification introduced with the presence of $T$ contributes to control this effect. In Fig.~\ref{fig6} we depict the energy density for $k=1$, $\alpha\!=\!1/2$ and $\beta$ within the interval \eqref{eq46}. In this case the split starts approximately at $\beta=0.012$.

\begin{figure}[h!]
\begin{center}
\includegraphics[scale=0.7]{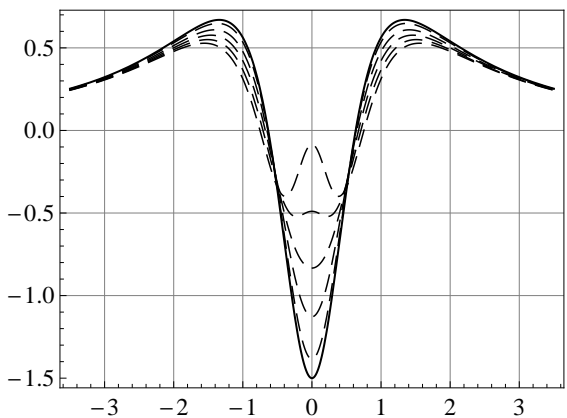}
\end{center}
\vspace{-0.7cm}
\caption{\small{The stability potential, depicted for $k=1$ and $\beta\!=\!0$ (solid line), $\beta\!=\!0.002, 0.006, 0.010, 0.014, 0.018$ (dashed lines).\label{fig7}}}
\end{figure}

The solutions of equation \eqref{eq45.a} can be written as
\begin{equation}\label{eq49}
\phi(y)=\pm \sqrt{\frac{6}{4\!-\!3\alpha}\!-\!\frac{320\beta k^2}{4-3\alpha}}~\mbox{EllipticE}~\Big(\phi_s; \frac{392\beta k^2}{160\beta k^2\!-\!3}\Big)\,,
\end{equation}
where $\phi_s=\arcsin\Big[\tanh(k y)\Big]$ and $\mbox{EllipticE}$ is the elliptic integral of the second kind. For $\beta=-3/232k^2$ and $\beta=0$ we obtain the solution as, respectively,
\bes\label{eq50}
\ben
\phi(y)&=& \sqrt{\frac{294}{116-87 \alpha}}~ \tanh (k y)\,,\label{eq50.a}\\
\phi(y)&=& \sqrt{\frac{6}{4-3 \alpha}} \arcsin\Big[\tanh (k y)\Big]\,.\label{eq50.b}
\een
\ees

We now study linear stability of the gravity sector. For this we make the following change of variables
\begin{equation}\label{eq51}
y=\frac1k \arcsinh (kz)\,.
\end{equation}
It allows to get the stability potential as
\ben
U(z)\!\!\!&=&\!\!\!-\frac{3k^2}4\frac{2-5k^2z^2}{(1+k^2 z^2)^2}+\frac{56\beta k^4(1-3k^2z^2)}{(1+k^2 z^2)^2 f(z)}-\nn
\!\!\!&&\!\!\!-\frac{3136 k^8 z^2 \beta^2}{(1 + k^2 z^2)^2f(z)^2}\,,
\een
where, $f(z)=1+k^2z^2-8\beta k^2(2-5k^2z^2)$.
This potential is depicted in Fig.~\ref{fig7}. The appearance of a lump in the well confirms the splitting of the brane as $\beta$ increases. We note that the stability potential does not depend on $\alpha$, although $\alpha$ contributes to split the brane, as shown above in \eqref{eq48}, for instance.

The results show that the presence of $R^2$ controls the splitting of the brane, as also noted in Ref.~\cite{Bazeia:2013uva}; here, however, the $T$ term also contributes, adding quantitative modification to the braneworld profile.  

We can also consider adding a quadratic term in $T$ with a quadratic term in the scalar curvature. In this case, we have
\begin{equation}
F=R+\beta R^2-\alpha T^2\,.
\end{equation}
The equations \eqref{eq9} become
\bes
\ben
&&-\frac23\phi^{\prime 2}+\alpha \phi^{\prime 2}T=\nn
&&A^{\prime\prime}
+\frac83\beta \Big(\!2 A^{\prime \prime \prime \prime }\!\!+\!16 A^{\prime\prime 2}\!\!+\!8A^\prime \!\!A^{\prime \prime \prime } \!\!+\!5 A^{\prime 2}\!\! A^{\prime \prime }\Big)\,,\\
&&\!\!\!\!\!\!V(\phi)\!-\!\frac12\phi^{\prime 2}+\!\frac{3\alpha }2 \phi^{\prime 2} T-\frac{\alpha}{4}T^2=\nn
&&-3A^{\prime 2}-4\beta\Big(5  A^{\prime 4}\!\!+8 A^{\prime }A^{\prime \prime \prime}\!\!+32  A^{\prime 2 }A^{\prime \prime }\!\!-4  A^{\prime \prime 2}\Big).\;\;\;\;
\een
\ees
To solve the model we assume that the warp function is yet given by \eqref{eq44}, and we take $k=1$. This problem is much harder then the previous one, but we then rewrite the above equations in the simpler form
\bes
\ben
&&\phi^{\prime 2}-\frac{3\alpha}{2} \phi^{\prime 2}T= \frac{3}2 S^2+4\beta \Big(29S^2-49S^4\Big),\\
&&V(\phi)\!-\!\frac12\phi^{\prime 2}+\!\frac{3\alpha }2 \phi^{\prime 2} T-\frac{\alpha}{4}T^2=\nn
&&\;\;\;\;\;\;\;\;\;\;-3+3S^2
-4\beta\Big(5  - 58 S^2 + 49 S^4\Big).
\een
\ees

\begin{figure}[t]
\begin{center}
\includegraphics[scale=0.7]{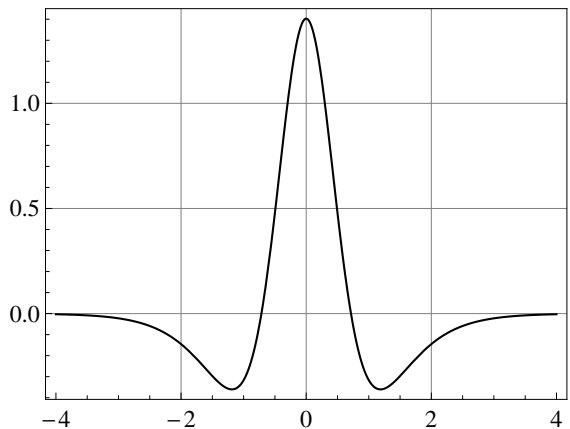}
\end{center}
\vspace{-0.7cm}
\caption{\small{The energy density, depicted for the solution \eqref{energy2}.\label{fig8}}}
\end{figure}

These equations have analytical solutions for appropriate choices of the parameters $\alpha$ and $\beta$. We have examined some specific cases, and they are all similar. Thus, we take, for instance, $\alpha=0.02304$ and $\beta=0.00433$; the solutions are
\bes
\ben
\phi(y)\!\!&=&\!\!\pm\, p \arcsin[\tanh(y)]\,,\\
V(\phi)\!\!&=&\!\!-2.31489+3.00269 \cos^2\Big(\frac{\phi}{p}\Big)\,,
\een
\ees
where $p=1.19576$. The energy density becomes
\begin{equation}\label{energy2}
\rho(y)=-2.31489\, S^2+3.71762\,  S^4\,.
\end{equation}
It is depicted in Fig~\ref{fig8}, showing standard profile. The absence of splitting is due to the fact that the analytical solution is obtained for a very small value of $\beta$.

\section{COMMENTS AND CONCLUSIONS}\label{sec:end} 

In this work we studied some thick braneworld scenarios in an $AdS_5$ warped geometry with a single extra dimension of infinite extent, with the gravity sector extended to accommodate the $F(R,T)$ modification, coupled to a single real scalar field with standard kinematics. We used the equations of motion to show explicitly that the gravity sector is linearly stable when we consider $F(R,T)=G(R)+H(T)$. 

We investigated two distinct models, the first one with $F(R,T)=R-\alpha\, T$, and the other with
$F(R,T)=R+\beta R^2-\alpha T$. In the first case, the contributions that appear from the extra term $T$ induces quantitative modifications in the thick brane profile, without changing its qualitative behavior. In the second case, the presence of $R^2$ works to split the brane. This effect was identified before in \cite{Bazeia:2013uva}, and the modification introduced with $\alpha \,T$ also contributes to the splitting, but it adds no new effect in the braneworld profile. We then conclude that the modification of the form $R\to F(R,T)=G(R)+H(T)$ does not destabilize the gravity sector of the thick braneworld with an $AdS_5$ warped geometry with a single extra dimension of infinite extent. Moreover, the presence of the term $H(T)=-\alpha\, T$ contributes to change the braneworld profile quantitatively, although it adds no new qualitative effect to the brane. If we add the term $H(T)=-\alpha T^2$, it seems that the effect is similar, but here, however, one needs to explore a larger range of parameters. This task requires numerical investigation, an issue that is out of the scope of the work.  

The present study drives our attention to some issues of current interest, one concerning the investigation of how the quantitative modifications that we have identified in this work act to localize fermion and gauge fields within the thick brane \cite{fl,gl}. Another, very natural continuation is related to the hybrid brane recently proposed, in which one changes the potential of the source field to make the scalar field solution compact \cite{we2}, to see how it adds in this new scenario. This last issue is now under consideration, and we hope to report on it in the near future.

\acknowledgements

DB and RM would like to thank CNPq for partial financial support.



\begin{thebibliography}{99}

\bb{RS}L. Randall and R. Sundrum, Phys. Rev. Lett. {\bf83}, 4690 (1999).
\bb{GW}W.D. Goldberger and M.B. Wise, Phys. Rev. Lett. {\bf83}, 4922 (1999).
\bb{F}O.~DeWolfe, D.~Z.~Freedman, S.~S.~Gubser and A.~Karch, Phys.\ Rev.\ D {\bf 62}, 046008 (2000).
\bb{C}C. Csaki, J. Erlich, T. Hollowood, Y. Shirman, Nucl. Phys. B {\bf581}, 309 (2000); C. Csaki, J. Erlich, G.
Grojean, T. Hollowood, Nucl. Phys. B {\bf584}, 359 (2000).
\bb{G} M. Gremm,  Phys.\ Lett.\ B {\bf 478}, 434 (2000);
F.A. Brito, M. Cvetic, and S.-C. Yoon, Phys. Rev. D {\bf64}, 064021 (2001);
M. Cvetic and N.D. Lambert, Phys. Lett. B {\bf540}, 301 (2002);
A. Campos, Phys. Rev. Lett. {\bf88}, 141602 (2002);
A. Melfo, N. Pantoja, and A. Skirzewski, Phys. Rev. D {\bf67}, 105003 (2003); D. Bazeia, F.A. Brito, and J.R. Nascimento,
Phys. Rev. D {\bf68}, 085007 (2003); 
D. Bazeia, C. Furtado, and A.R. Gomes, JCAP {\bf0402}, 002 (2004);
 D. Bazeia and A.R. Gomes, JHEP {\bf0405}, 012 (2004).
\bb{ghost}C.M. Will, Living Rev. Rel. {\bf9}, 3 (2006); 
R.P. Woodard, Lect. Notes Phys. {\bf720}, 403 (2007).
\bb{psv}P. Pani, T.P. Sotiriou and D. Vernieri, Phys.\ Rev. D {\bf88}, 121502 (2013).
\bb{FR}T.P. Sotiriou and V. Faraoni, Rev. Mod. Phys. {\bf82} 451 (2010);
A. De Felice, S. Tsujikawa, Living Rev. Rel. {\bf13}, 3 (2010). 
\bb{c1}T. Harko, F. S. N. Lobo and E. N. Saridakis, {\it Cosmology with higher-derivative matter fields.} arXiv:1405.7019.
\bb{c2}D. Bazeia, F.A. Brito, and F.G. Costa, Phys. Rev. D {\bf90}, 043523 (2014).
\bb{gly}Bin Guo, Yu-Xiao Liu, and Ke Yang, {\it Brane worlds in gravity with auxiliary fields}. arXiv:1405.0074.
\bb{we}D. Bazeia, A.S. Lob\~ao Jr., and R. Menezes, Phys. Rev. D {\bf90} (2014); D. Bazeia, M.A. Marques, R. Menezes, and D.C. Moreira, {\it New braneworld models in the presence of auxiliary fields}, arXiv:1412.XXXX.
\bb{m1}Yu-Xiao Liu, Yuan Zhong, Zhen-Hua Zhao, Hai-Tao Li, JHEP {\bf1106}, 135 (2011).
\bb{m2}Yi Zhong, Yu-Xiao Liu, Feng-Wei Chen, Qun-Ying Xie, {\it Warped Brane worlds in Critical Gravity}, arXiv:1403.5109;
Zeng-Guang Xu, Yu-Xiao Liu, Yuan Zhong, {\it Metastable gravitons in $f(R)$-brane models}, arXiv:1405.6277; D. Bazeia, L. Losano, R. Menezes, G.J. Olmo, and D. Rubiera-Garcia, {\it Thick brane in $f(R)$ gravity with Palatini dynamics,} arXiv:1411.0897.
\bb{RT1}T.~Harko, F.~S.~N.~Lobo, S.~Nojiri and S.~D.~Odintsov, Phys.\ Rev.\ D {\bf 84}, 024020 (2011).
\bibitem{RT2}F.~G.~Alvarenga, A.~de la Cruz-Dombriz, M.~J.~S.~Houndjo, M.~E.~Rodrigues and D.~S\'aez-G\'omez,
Phys. Rev. D {\bf87}, 103526 (2013).
\bb{Bazeia:2013uva}D.~Bazeia, A.~S.~Lob\~ao Jr., R.~Menezes, A.~Y.~Petrov and A.~J.~da Silva, Phys. Lett. B {\bf729}, 127 (2014).
\bb{fl}R.A.C. Correa, A. de Souza Dutra, and M.B. Hott, Class. Quant. Grav. {\bf28}, 155012 (2011); A.E.R. Chumbes, A.E.O. Vasquez,
and M.B. Hott, Phys. Rev. D {\bf83}, 105010 (2011); Yu-Xiao Liu, Zeng-Guang Xu, Feng-Wei Chen, and Shao-Wen Wei, Phys. Rev. D {\bf89}, 086001 (2014).
\bb{gl}W.T. Cruz, A.R.P. Lima, C.A.S. Almeida, Phys. Rev. D, {\bf87}, 045018 (2013); W.T. Cruz, R.V. Maluf, C.A.S. Almeida, Eur. Phys. J. C {\bf73}, 2423 (2013).
\bb{we2}D. Bazeia, L. Losano, M.A. Marques, and R. Menezes, EPL {\bf107}, 61001 (2014); Phys. Lett. B {\bf736}, 515 (2014). 
\end{thebibliography}
\end{document}